\documentclass[12pt]{article}
\usepackage{amsmath,amsthm,amsfonts}
\usepackage{cite}
\theoremstyle{definition}

\theoremstyle{definition}

\theoremstyle{definition}
 
\begin{document}
\title{\LARGE 
A class of hermitian generalized Jordan \\
triple systems and Chern-Simons gauge theory\\}
\date{}
\author{\Large Noriaki Kamiya \footnote
           {
e-mail address : kamiya@u-aizu.ac.jp}\\
\it Department of Mathematics, University of Aizu\\
\it Aizuwakamatsu, 965-8580, Japan\\ 
\Large Matsuo Sato \footnote
           {
e-mail address : msato@cc.hirosaki-u.ac.jp}\\
\it Department of Natural Science, Faculty of Education, Hirosaki University\\ 
\it Bunkyo-cho 1, Hirosaki, Aomori 036-8560, Japan\\}

\maketitle
\thispagestyle{empty}

\begin{abstract}
We find a class of hermitian generalized Jordan triple systems (HGJTSs) and hermitian $(\epsilon, \delta)$-Freudenthal-Kantor triple systems (HFKTSs). We apply one of the most simple HGJTSs which we find to a field theory, and obtain a typical $u(N)$ Chern-Simons gauge theory with a fundamental matter.
\vskip 0mm
\noindent
APS classification: 11.15.Yc; 11.25.Yb; 11.25.-w 
\noindent
\end{abstract}
\vskip 3mm
{\bf Introduction }
\vskip 2mm
\par
HGJTSs and HFKTSs \cite{KamiyaSato1, KamiyaSato2} are generalizations\footnote{
A different generalization, the so-called differential crossed module, was extensively investigated in \cite{modules1, modules2, modules3}. } of hermitian 3-algebras \cite{Nambu, Filippov, BergshoeffSezginTownsend, deWHN, Yoneya, Minic, BLG1, BLG2, BLG3, BLG4, BLG5, Nogo1, Nogo2, Nogo3, sp1, Iso, sp2, sp3, Lorentz0, Nogo4, Nogo5, Nogo6, Nogo7, text, Nogo8, Nogo9, Nogo10, NishinoRajpoot, GustavssonRey, ABJ, HanadaMannelliMatsuo, IshikiShimasakiTsuchiya, KawaiShimasakiTsuchiya, IshikiShimasakiTsuchiya2, PangWang, DeBellisSaemannSzabo, Jabbari, GomisSalimPasserini, HosomichiLeeLee, Smolin1, Smolin2, Azuma, 3-graded, 5-graded, Palmkvist}, which have played crucial roles in M-theory. The field theories applied with hermitian 3-algebras are the Chern-Simons gauge theories with bi-fundamental matters that describe the effective actions of membranes in M-theory\footnote{The structures of generalized Jordan triple systems were found in the effective actions of the membranes in \cite{Nilsson, Chowdhury}}.  Moreover, 3-algebra models of M-theory itself have been proposed and were studied in \cite{bosonicM, MModel, LorentzianM, Zariski, book, 3BFSS, 3IIB, 4IIB}. Therefore, it is interesting to apply  HGJTSs and HFKTSs to field theories.  
\par
In \cite{KamiyaSato1, KamiyaSato2}, we searched HGJTSs and HFKTSs by adding some terms to the Hermitian 3-algebras. In this paper, we study a class of  HGJTSs and HFKTSs systematically. We apply one of the most simple HGJTSs which we find to a field theory, and obtain a typical $u(N)$ Chern-Simons gauge theory with a fundamental matter. 
\par
Here, we are concerned with algebras and
triple systems that are finite or infinite over a 
complex number field, unless otherwise specified. This paper is an announcement of new results, that is first step of our works for hermitian triple systems and applications to field theories. hence we will give a lot of references \cite{3,4,5,6,7,8,9,10,11,12,13,14,15,16,17}. 
We refer to \cite{1, 2, 18, 19, 20} for nonassociative algebras, and to \cite{21, 22, 23} for Lie superalgebras, for example.
\par
\section{Definitions}
\par
In this section, we summarize definitions of HGJTS and HFKTS given in \cite{KamiyaSato1, KamiyaSato2}.

 A triple system $U$ is a vector space over a field $F$ of characteristic $\neq 2, 3$ with a trilinear map $U \otimes U \otimes U \to U$. In this paper, we are concerned with triple systems over the complex number $\bold{C}$, and denote the trilinear product by $<xyz> \in U$ for $x,y,z \in U$, assuming that $<xyz>$ is $\bold{C}$-linear on $x$ and $z$, and $\bold{C}$-antilinear on $y$. Also we use the notations of two operations $L(x,y) \in$ End $U$ and $K(x,y)\in$ End $U$ with  respect to the triple systems ([1], [2]), where $L(x,y)z=<xyz>$ and $K(x,y)z=<xzy>- \delta <yzx>\ \  (\delta =\pm 1)$.
\vskip 3mm
{\bf Definition.}
\quad
A triple system
$U$ is said to be a *-$(\epsilon, \delta)$-Freudenthal-Kantor triple
system (FKTS) if relations (0)--(iv) are satisfied.
\par
0)\quad  $U$ is a Banach space,
\par
$
i) \ [L(x,y),L(z,w)]=L(<xyz>,w)+\varepsilon L(z,<yxw>) \quad\mbox{(fundamental identity)},
$
\par
\noindent
where $[A,B]=AB-BA$, $A, B \in \mbox{End}\; U, and \ \ \varepsilon =\pm 1$,
\par
$
ii) \ K(<xyz>,w)+K(z,<xyw>)+\delta K(x,K(z,w)y)=0 \quad\mbox{(second order identity)},
$
\par
iii)\quad $<xyz>$ is a ${\bf C}$-linear operator on
$x$ and $z$, and a ${\bf C}$-antilinear operator on $y$,
\par
iv)\quad $<xyz>$ is continuous with respect to a norm,
$||\ ||$ 
that is,
there exists 
\par
\noindent
$K>0$ such that
$$
||<xxx>||\leq K || x ||^{3}
\ {\rm for\ all} \; x\in U.$$
$U$ is called *-generalized Jordan triple system (GJTS) if 0), i), iii), and iv) are satisfied when $\epsilon=-1$ and $\delta=-1$. The condition ii) is called second order identity and *-generalized Jordan triple system of second order is (-1, -1) FKTS. Note that there are many simple Lie algebras (the case of $\delta=1$) and simple Lie superalgebras (the case of $\delta=-1$) constructed from $(\epsilon, \delta)$-Freudenthal-Kantor triple systems ([4,7,8]) because of the second order identity.
\par
Furthermore, a *-$(\epsilon, \delta)$-FKTS  and *-GJTS are said to
be hermitian if
they satisfy the following condition.
\par
v)\quad 
All operators $L(x,y)$
are operators with
a hermitian metric
$$
(x,y)=tr \ L(x,y),$$
that
is,
$(L(x,y)z,w)=(z,L(y,x)w)$
and
$(x,y)=\overline{(y,x)}$.
\par
\vskip 3mm

\section{A class of HGJTSs and HFKTSs}
In this section, we obtain a class of HGJTSs and HFKTSs by making an ansatz $<xyz>=\alpha x \tilde{y} z + \beta z \tilde{y} x + \gamma \bar{y} \tilde{\bar{x}} z$ and explicitly evaluating the conditions i) and ii) in the last section. The hermitian 3-algebra $x \tilde{y} z-z \tilde{y}x$ is in a special case when $\alpha=1$, $\beta=-1$, and $\gamma=0$. By straightforward, but extensive calculations (to be omitted the proof here), we obtain theorems as follows.

\par
\vskip 3mm
{\bf Theorem A.}
\quad
Let $U$ be a hermitian triple system over complex number $\bold{C}$ satisfying the triple product
$<xyz>=\alpha x \tilde{y} z + \beta z \tilde{y} x + \gamma \bar{y} \tilde{\bar{x}} z$ for $\alpha, \beta, \gamma \in \bold{R}$ (real number), $x, y, z \in U$, induced from an associative algebra with an involution $\widetilde{xy} = \tilde{y} \tilde{x}$ and $\tilde{\tilde{x}}=x.$ 

Then we have a necessary and sufficient condition for $U$ to be a hermitian generalized Jordan triple system is that the following hold;

$\gamma( \alpha + \gamma)=0.$

Furthermore, we have for $U$ to have the second order condition ii) is the following holds;

$\alpha=\beta$ if $\delta=1$,

 $\alpha=-\beta$ if $\delta=-1$.

\par
\vskip 3mm

Summarizing, we obtain a class of HGJTSs,
\par
\vskip 3mm
{\bf Proposition B.}
\quad
Let $U$ be a set of $M(n, n; \bold{C})$ with the triple product defined by 

1)$<xyz>$ is a linear combination of $x \tilde{y} z$ and $z \tilde{y}x$,

2) $<xyz>$ is a linear combination of $x \tilde{y} z-\bar{y} \tilde{\bar{x}} z$ and $z \tilde{y}x$,

where $\tilde{x}=\bar{x}^T$, $\bar{}$ is the conjugation of $\bold{C}$, and $x^T$ is the transpose of $x$. Then $U$ are HGJTSs. The hermitian 3-algebra $<xyz>= x \tilde{y} z-z \tilde{y}x$ is in a special case of 1). One of the most simple cases, $<xyz>=x \tilde{y} z$ will be applied to a field theory in the next section.

\par
\vskip 3mm

We also obtain a class of HFKTSs,
\par
\vskip 3mm
{\bf Proposition C.}
\quad
Let $U$ be a set of $M(n, n; \bold{C})$ with the triple product defined by 

a) $<xyz>=x \tilde{y} z + z \tilde{y} x$,

b) $<xyz>=x \tilde{y} z + z \tilde{y} x -\bar{y} \tilde{\bar{x}} z$,

c) $<xyz>=x \tilde{y} z - z \tilde{y} x$,

d) $<xyz>=x \tilde{y} z - z \tilde{y} x +\bar{y} \tilde{\bar{x}} z$,

where $\tilde{x}=\bar{x}^T$, $\bar{}$ is the conjugation of $\bold{C}$, and $x^T$ is the transpose of $x$. Then we have the case of a) and b) are (-1,1)-HFKTSs and the case of c) and d) are (-1,-1)-HFKTSs.

\par
\vskip 3mm
{\bf Remark. }
\quad
For the case of nonassociative and nonhermitian algebras, we note that there is a structureable algebra due to Allison \cite{1} satisfying the triple product
$<xyz>=(x \tilde{y})z+(z\tilde{y})x-(z\tilde{x})y$ and $\widetilde{xy}=\tilde{y}\tilde{x}$, and $\tilde{\tilde{x}}=x$, which is a (-1,1)-Freudenthal-Kantor triple system also this algebra contains a class of commutative Jordan algebras and associative algebras.

We call a hermitian ($\alpha$, $\beta$, $\gamma$) structureable algebra about the hermitian triple system induced from a nonassociative algebra satisfying 
$<xyz>=\alpha (x \tilde{y})z +\beta (z\tilde{y})x+ \gamma(z\tilde{\bar{x}})\bar{y}$. However, the details will be discussed in other paper.

\section{Chern-Simons gauge theory with a fundamental matter}
\par
In this section, we apply one of the most simple HGJTSs to a field theory. We define a 3-bracket, $[x, \overline{y}, z]:=<x y z>$.

We start with
\begin{eqnarray}
S= \int d^3x \mbox{tr}(\! \! \! \!\! \! \! \!\! &&-\bold{D}_{\mu} Z \overline{\bold{D}^{\mu} Z}^T + V(Z) \nonumber \\
 &&+\epsilon^{\mu\nu\lambda}(-\frac{1}{2}A_{\mu \bar{b}c}\partial_{\nu} A_{\lambda \bar{d}a} \overline{T}^{T\bar{d}}[T^c, \overline{T}^{\bar{b}}, T^a ] \nonumber \\
&& \qquad \quad +\frac{1}{3} A_{\mu \bar{d} a} A_{\nu \bar{b} c} A_{\lambda \bar{f} e}[T^c, \overline{T}^{\bar{b}}, T^a] \overline{[T^f, \overline{T}^{\bar{e}}, T^d]})), \label{masteraction}
\end{eqnarray}
where
\begin{equation}
\bold{D}_{\mu} Z=\partial_{\mu} Z -A_{\mu \bar{b} a} [T^a, \overline{T}^{\bar{b}}, Z].
\end{equation}
$Z$ and $A_{\mu}$ are matter and gauge fields, respectively. $V(Z)$ is a potential term. This action is invariant under the transformations generated by the operator $L(x,y)-L(y,x)$. The action (\ref{masteraction}) equipped with  a Lorentzian Lie 3-algebra and a hermitian 3-algebra describes the bosonic parts of the effective actions of supermembranes in M-theory. We apply a HGJTS, 
$[x, \overline{y}, z]=<x y z>=x{\bar y}^{T}z$ in the previous section 
to this action and see what kind of field theory we obtain.

The covariant derivative is explicitly written down as

\begin{equation}
\bold{D}_{\mu} Z= \partial_{\mu} Z -iA_{\mu} Z,\end{equation}
where $A_{\mu}:= -i A_{\mu \bar{b}a} T^a \overline{T}^{T \bar{b}}$ are hermitian matrices, which generate the $u(N)$ Lie algebra. In order to see how gauge transformations act on these fields, let us see how $L(x,y)-L(y,x)$ acts on the coupling between $A_{\mu}$ and $Z$:
\begin{eqnarray}
&&\delta (A_{\mu \bar{b} a}[T^a, \overline{T}^{\bar{b}}, Z]) \nonumber \\
&=& \Lambda_{\bar{d} c} A_{\mu \bar{b} a }([[T^c, \overline{T}^{\bar{d}}, T^a], \overline{T}^{\bar{b}} Z]
-[T^a, \overline{[T^d, \overline{T}^{\bar{c}}, T^b]}, Z]
+[T^a, \overline{T}^{\bar{b}}, [T^c, \overline{T}^{\bar{d}}, Z]] \nonumber \\
&=& 
[i \Lambda, i A_{\mu}]Z
+i A_{\mu}i \Lambda Z \nonumber \\
&=&  i \Lambda i A_{\mu}Z,
\end{eqnarray}
where gauge parameters $\Lambda$ is defined in the same way as  $A_{\mu}$. From this, we obtain gauge transformations,
\begin{eqnarray}
\delta A_{\mu} &=&[i \Lambda, A_{\mu}] \nonumber \\
\delta Z &=& i \Lambda Z.
\end{eqnarray}
Therefore, $A_{\mu}$ transforms as adjoint representations of $u(N)$, whereas $Z$ transforms as a fundamental representation of $u(N)$. As a result, the action can be rewritten in a covariant form and we obtain a typical $u(N)$ Chern-Simons gauge theory with a fundamental matter,
\begin{eqnarray}
S&=&\int d^3x \mbox{tr} (-(\partial_{\mu} Z -iA_{\mu} Z) \overline{(\partial_{\mu} Z -iA_{\mu} Z)}^T +V(Z)\nonumber \\
&& + \epsilon^{\mu\nu\lambda}(\frac{1}{2}A_{\mu} \partial_{\nu} A_{\lambda}
+\frac{i}{3}A_{\mu} A_{\nu} A_{\lambda}). \label{CSaction}
\end{eqnarray}

\vspace*{1cm}

\section*{Acknowledgements}

The work of M.S. is supported in part by Grant-in-Aid for Young Scientists (B) No. 25800122 from JSPS.

\end{document}